\documentclass{article}

\usepackage{arxiv}

\usepackage[utf8]{inputenc} 
\usepackage[T1]{fontenc}    
\usepackage{hyperref}       
\usepackage{url}            
\usepackage{booktabs}       
\usepackage{amsfonts}       
\usepackage{nicefrac}       
\usepackage{microtype}      
\usepackage{lipsum}
\usepackage[export]{adjustbox}

\usepackage[section]{placeins}
\usepackage[shortlabels]{enumitem}
\usepackage{subfig}
\usepackage[numbers]{natbib}
\usepackage{amsmath,amssymb,amsfonts}
\usepackage{algorithmic}
\usepackage{graphicx}
\usepackage{textcomp}
\usepackage{xcolor}
\def\BibTeX{{\rm B\kern-.05em{\sc i\kern-.025em b}\kern-.08em
    T\kern-.1667em\lower.7ex\hbox{E}\kern-.125emX}}
    
\usepackage{verbatim}  

\usepackage{hyperref}
\usepackage{csquotes}
\usepackage{todonotes}
\usepackage{float}

\usepackage{diagbox}
\usepackage{multirow}
\usepackage{tabularx}
\newcolumntype{Y}{>{\centering\arraybackslash}X}

\usepackage{ tipa }

\usepackage[utf8]{inputenc}

\title{Assessing the robustness of critical behavior in stochastic cellular automata}

\author{Sidney Pontes-Filho$^{a,b,*}$, Pedro Lind$^{a,c,d}$ and Stefano Nichele$^{a,c,d,e,f}$
\bigskip \\ 
$^a$Department of Computer Science, Oslo Metropolitan University, Oslo, Norway\\
$^b$Department of Computer Science, Norwegian University of Science and Technology, Trondheim, Norway\\
$^c${\it AI Lab} -- OsloMet Artificial Intelligence Lab, Oslo, Norway\\
$^d${\it NordSTAR} -- Nordic Center for Sustainable and Trustworthy AI Research, Oslo, Norway\\
$^e$Department of Holistic Systems, Simula Metropolitan Centre for Digital Engineering, Oslo, Norway\\
$^f$Department of Computer Science and Communication, Østfold University College, Halden, Norway\\
$^*$Corresponding author: sidneyp@oslomet.no}

\date{\small}

\begin{document}
\maketitle
\begin{abstract}
There is evidence that biological systems, such as the brain, work at a critical regime robust to noise, and are therefore able to remain in it under perturbations. In this work, we address the question of robustness of critical systems to noise. In particular, we investigate the robustness of stochastic cellular automata (CAs) at criticality. A stochastic CA is one of the simplest stochastic models showing criticality. The transition state of stochastic CA is defined through a set of probabilities. We systematically perturb the probabilities of an optimal stochastic CA known to produce critical behavior, and we report that such a CA is able to remain in a critical regime up to a certain degree of noise. We present the results using error metrics of the resulting power-law fitting, such as Kolmogorov-Smirnov statistic and Kullback-Leibler divergence. We discuss the implication of our results in regards to future realization of brain-inspired artificial intelligence systems.
\end{abstract}

\section{Introduction}
\label{sec:intro}

Critical phenomena in general share two important fea\-tu\-res \citep{christensenbook}. First, they show an infinite correlation length, which means that information spans throughout several 
scales, both in time and in space. 
Second, they occur between two well-defined phases each one assuming
a specific range of values of an observable which can be "tuned" or controlled by one or more parameters.
These parameters, called control parameters, can drive the system to switch between phases and for specific values, so-called critical values, one observes critical behavior.
While most critical phenomena share these two ingredients, the second one is sometimes not
observed, which means the system still shows correlation spanning through different scales, but
there is no explicit control parameter to be tuned: independently of how we tune the system and
initialize it, the system always evolves {\it towards} the critical state. In other words,
the critical state is a stable state, an attractor of the system. This sort of critical
behavior is called self-organized criticality (SOC), which is one of the most striking non-linear phenomena found in nature.
Since its discovery in the 80s \citep{bak1987self,bak1988self}, several natural phenomena have been reported as showing SOC, ranging from the stock market \citep{mandelbrodtbook,joao2013,joao2012} to the brain \citep{heiney2021criticality,fontenele2019}.

Indeed, brain functioning involves the coordination of neural activity across several scales,
ranging from few neurons to large brain neural networks, leading to a natural resemblance to 
critical phenomena \citep{cocchi2017}, and since this functioning is the "natural" state of
the brain, it may be reasonable to hypothesize that this criticality follows some principles
of self-organization.
Recent discussions and investigations point indeed towards the possibility that SOC is the main factor for intelligence in the human brain \citep{heiney2021criticality} and therefore such findings from physics of critical phenomena
may help to investigate how to learn from brain dynamics in order to improve the capacity 
of computation of artificial intelligence (AI) systems.
How can self-organized criticality emerge in simple computational systems?

In this work, we address this question using cellular automata (CAs). 
CAs comprehend a family of models, in which a set of elementary "cells" form a lattice,
typically with one or two dimensions \citep{wolframbook}. 
The lattice iteratively evolves, and each cell takes
one of a countable number of states. CAs are, therefore, models with discrete
space, discrete time, and discrete state space.
\begin{figure*}[t]
\centering
\subfloat[Optimal stochastic CA]{\label{fig:ca_0.0}\includegraphics[width=0.45\textwidth]{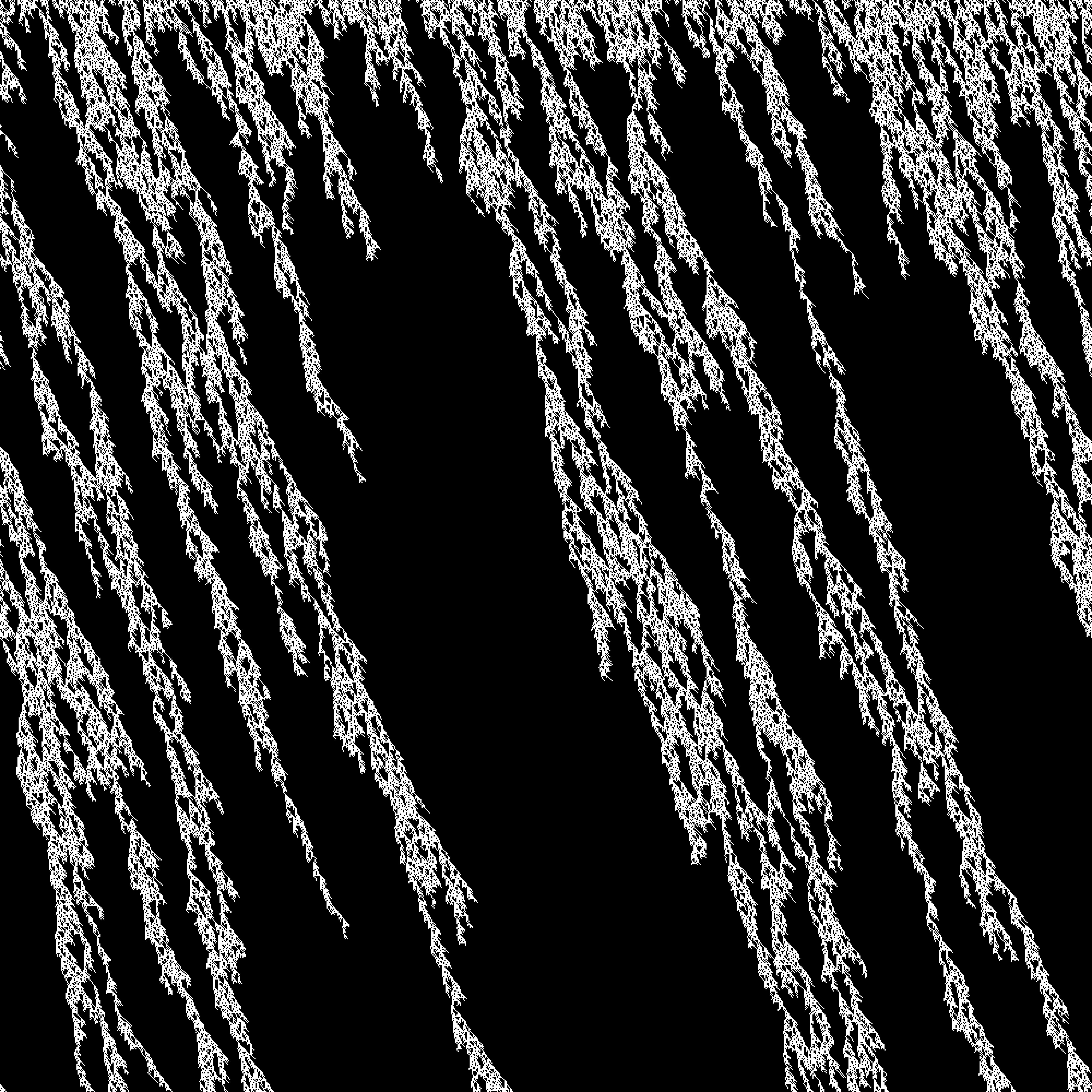}}%
\hspace{0.5cm}
\subfloat[Avalanche definition]{\label{fig:ava_definition}\includegraphics[width=0.45\textwidth]{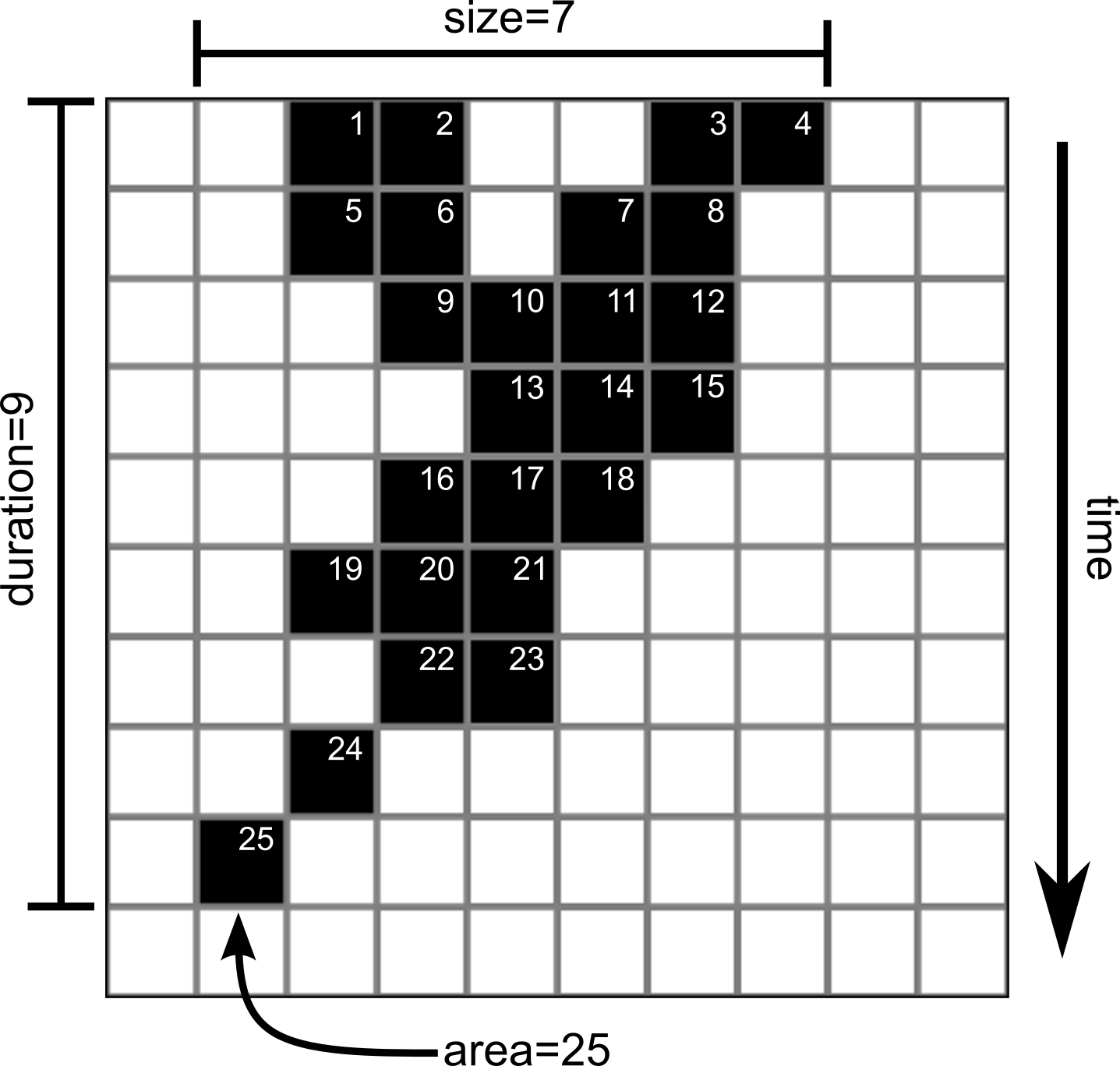}}%
\caption{
    \protect\subref{fig:ca_0.0} Illustration of the evolution of an optimal stochastic CA for critical behavior. The CA started with $N=1,000$ lattice squares and its evolution is plotted for $T=1,000$ time steps from top to bottom. The number of avalanches and the distributions of this illustration is shown in Fig.~\ref{fig:pertubation} (first row). \protect\subref{fig:ava_definition} Illustration of the three properties characterizing an avalanche: its area, defined by the total number of active sites in the same connected set; its size, given by the total number of different cells (sites) belonging to the avalanche, at least at one iteration; its duration, given by the total number of (successive) iterations that include at least one site in the avalanche.
    In this example, we have size of 7 cells, a duration of 9 iterations, and a total area of 25 sites.
    Active cells are black and have state 1. The boundary of the CA also limits an avalanche.}
\label{fig01}
\end{figure*}

Due to their simple implementation, they have been used to approach several complex phenomena, namely those showing large-scale correlations as a result of short-range, typically nearest neighbor, interactions \citep{vichniac1984}.
The update of each cell considers the composition of its state together with
the state of its nearest neighbors.
From the very beginning, studies on CAs have also reported the emergence of critical behavior in general \citep{chate1990criticality,singha2020chimera} and SOC in particular \citep{bak1989self},
where a prototypical example is the so-called Game of Life \citep{berlekamp2004winning}.
Recently, the concept of SOC in Game of Life was used in the context of machine behavior to introduce collective robots with simple control and local interactions \citep{rahwan2019machine},
as well as for discussing the general features of artificial life \citep{gershenson2020self}.

The rules governing the evolution of each cell and its coupling with nearest neighbors can be deterministic: state configurations of one cell and its neighboring cells impose always the same state on the cell in the next iteration.
A more realistic extension of such a model is to enable the iteration throughout CA's evolution to be updated according to some rules, with a certain {\it probability} $p$ not necessarily 0 or
1. Such CAs are usually called stochastic CAs \citep{pontes2020neuro,langton1990computation}.
An example of a one-dimensional stochastic CA is shown in Fig.~\ref{fig:ca_0.0}.
Indeed, if the brain functioning shows critical behavior, it should involve critical states which can be achieved through dynamical processes driven by some stochastic freedom. Stochastic CAs seem therefore to be a better choice to explore criticality in the context of artificial intelligence.

Recently \citep{pontes2020neuro}, using a genetic algorithm (please see \citep{holland1992genetic} for an introduction to genetic algorithms), the authors found eight {\it optimal} probabilities that grant critical behavior in a 1D stochastic CA with 3 neighbors. 
Due to its freedom, this stochastic CA can have different outcomes, starting from the same initial configuration of the cells composing the CA, all of them showing critical behavior. 
One question which remains unanswered is how {\it sensitive} this stochastic CA is to
small changes of the optimal probabilities.
The brain functioning is driven by dynamical processes with some stochasticity, but it
is also robust against changes in its stochastic features. Namely, its critical behavior
is observed even under changes in the stochastic dynamics.
The goal of this work is to assess the {\it robustness} of the critical behavior in the {\it optimal} stochastic CA presented in \citep{pontes2020neuro}. 

We start in Sec.~\ref{sec:methods} by introducing the main tools and methods,
namely, how the {\it optimal} stochastic CA is obtained, the tools to uncover the
critical behavior in the evolution of one particular CA and the measures to quantitatively 
assess the robustness of its critical behavior.
In Sec.~\ref{sec:results}, we present our main results, reporting a broad range of probabilities
for which criticality remains in the evolution of the CA.
Sec. \ref{sec:conclusions} concludes the paper.

\section{Methods}
\label{sec:methods}

\subsection{Background and Prior Model}

A CA that is one-dimensional, 2-state, and based on local interactions evolves according to an update rule of the form
\begin{equation}
c_{i,t+1}={\cal F}(c_{i-1,t},c_{i,t},c_{i+1,t}),
\label{eq_neigh}
\end{equation}
where $c_{i,t}$ is either $0$ or $1$, denoting the state of cell $i$ in iteration $t$. 
Periodic boundary conditions, $c_{0,t}=c_{N,t}$ and $c_{N+1,t}=c_{1,t}$, are used where $N$ is the number of cells in the CA.
In general, the function ${\cal F}$ maps each of the 8 possible 3-tuple at iteration $t$ into the updated state of the middle cell $i$ in the next iteration $t+1$. 
If the CA is deterministic, there are exactly $2^8=256$ possible choices of ${\cal F}$. 

If the CA is stochastic, instead of such ${\cal F}$-functions, we define a function
${\cal P}$ which gives the probability for the state at cell $i$ in iteration
$t+1$ to be {\it one}, given the present state of the 3-tuple defining its nearest
neighborhood:
\begin{equation}
{\cal P}(c_{i-1,t},c_{i,t},c_{i+1,t}) = 
 \hbox{Pro} \Big ( \big [c_{i,t+1}=1 \big ] \vert c_{i-1,t},c_{i,t},c_{i+1,t}  \Big ).
\label{eq_prob}
\end{equation}
Function ${\cal P}$ is fully described by a vector of eight probabilities, one
for \ each \ possible \ 3-tuple $(c_{i-1,t},c_{i,t},c_{i+1,t})$.
\begin{table}[t]
\renewcommand{\arraystretch}{1.1}
\centering
\begin{tabular}{|c|c|} 
\hline
State at $t$ & \textbf{${\cal P}$} (Eq.~(\ref{eq_prob})) \\ \hline
(0,0,0) & 0.394221 \\ \hline
(0,0,1) & 0.094721 \\ \hline
(0,1,0) & 0.239492 \\ \hline
(0,1,1) & 0.408455 \\ \hline
(1,0,0) & 0.000000 \\ \hline
(1,0,1) & 0.730203 \\ \hline
(1,1,0) & 0.915034 \\ \hline
(1,1,1) & 1.000000 \\ \hline
\end{tabular}
\caption{Selected stochastic CA in \citep{pontes2020neuro}.}
\label{tab01}
\end{table}

In Fig.~\ref{fig:ca_0.0}, we present one example of a stochastic CA with periodic boundary condition, states uniformly initialized, and state transition probabilities shown in Tab.~\ref{tab01}. 
This choice of probabilities
{\it maximizes} the critical behavior of the CA in such a configuration of boundary condition and initialization. There is a dependency in this configuration because if the CA starts with too many cells in the strong quiescent state (state 1), the resting state can be reached faster, then requiring re-initialization because all the cells have the same state and become static. With the re-initialization, the CA will maintain its activity and more avalanches can be produced. Also, some boundary conditions can make the CA never reach the resting state.
The criticality of the CA will be assessed by measuring the statistical distribution
of features characterizing an avalanche in the CA evolution.
We define an avalanche during the evolution of a specific CA as the connected cluster
of active states - i.e.~state-1 cells - throughout the chain of cells composing the CA
and throughout its time evolution. Fig.~\ref{fig:ava_definition} illustrates an avalanche together
with its three main features. The area of the avalanche corresponds to the total number of active states throughout the CA and during the full evolution which forms a connected set.
The avalanche size is given by the total number of adjacent cells which are activated
($c=1$) at least once, during the avalanche.
The avalanche duration is given by the total number of successive iterations from the beginning 
until the end of the avalanche, including at least one site in the avalanche.
The boundary of the CA is also the end of an avalanche.
In the example of Fig.~\ref{fig:ava_definition} the avalanche has a size of 7 cells, a duration
of 9 iterations, and a total area of 25 sites.
Our tentative was to have a CA as wide and lasting as possible because we could get a broader range of avalanche size, duration and area. However, due to memory and time limitations, our selected number of lattice squares is $N=1,000$ and the number of time steps is $T=1,000$.

By keeping track of the avalanches emerging during the evolution of a stochastic
CA, we can record the distribution of their area, size and duration.
These three types of avalanche measurements are more similar to neuronal avalanches \citep{heiney2021criticality}.
Moreover, the closer the distribution is from a power-law the closer the
CA behavior is from a critical regime.
Therefore, we recently introduced a fitness score (see definition below), to assess how "critical" a particular
CA realization is \citep{pontes2020neuro}, based on different measures of the error
associated with the fit.
Using a genetic algorithm, we found
the set of eight probabilities that leads to the most critical CA realizations
(see
Tab.~\ref{tab01}).


Recently we implemented a Python library called {\it EvoDynamic}, to simulate several dynamical systems \citep{pontes2020neuro}, such as cellular automata, random Boolean networks, and echo state networks. 
This library is based on the {\it TensorFlow} deep neural network framework \citep{abadi2016tensorflow}, and it includes
genetic algorithms that can be used to evolve dynamical systems towards a desired behavior or dynamics.

EvoDynamic has a general and single implementation that can simulate various dynamical systems because they are, in essence, dynamical graphs or networks. 
Since the weighted adjacency matrix and mapping function are specific to a dynamical system, they need to be adapted to simulate a stochastic elementary cellular automaton because the implementation of EvoDynamic is based on artificial neural networks. 
To make this possible, the implementation of an artificial neural network is generalized 
to incorporate the possibility of simulating CA.
More specifically, EvoDynamic uses the general form of a feed-forward artificial neural network 
(ANN) without bias, $\mathbf{l}_{i+1} = a(\mathbf{W}\cdot \mathbf{l}_{i})$,
with $\mathbf{l}_{i}$, the layer index $i$ of the ANN; $\mathbf{W}$, a weight matrix; and 
$a$, an activation function, and interprets it as a non-linear dynamical system with 
discrete time steps, namely:
\begin{equation}
\label{eq_general}
\mathbf{c}_{t+1} = f(\mathbf{A}\cdot \mathbf{c}_{t}).
\end{equation}
where $\mathbf{c}_{t}$ is a group of cells (or nodes in a graph) in iteration $t$, $\mathbf{A}$ 
is the weighted adjacency matrix, and $f$ represents the (non-linear) mapping function. For the stochastic CA, the weighted adjacency matrix $\mathbf{A}$ connects the center cell $c_{i,t}$ with its three neighbors $(c_{i-1,t},c_{i,t},c_{i+1,t})$. To identify which neighbor has state 0 and 1 after the matrix multiplication between $\mathbf{A}$ and $\mathbf{c}_{t}$, the weights assigned to $(c_{i-1,t},c_{i,t},c_{i+1,t})$ are $(4,2,1)$. Therefore, each of the eight neighborhood combinations is represented by a unique number from 0 to 7, which the mapping function $f$ maps it to its corresponding probability for the random generation of state 0 or 1 in the next iteration.

The EvoDynamic framework was used to find the eight probabilities in Tab.~\ref{tab01} which maximizes the critical behavior of a stochastic CA, applying a so-called genetic algorithm.
The genetic algorithm, in general, mimics a "natural selection" process, searching for optimal parameter values (the probabilities) by changing the parameter values and maximizing a pre-defined
fitness function that calculates a fitness score.
Instead of maximizing the function through a "supervised" path, such as a gradient
descendent scheme, the genetic algorithm updates the values of the probabilities, starting from some set of initial values, and then applying random perturbations. If the perturbation increases the fitness
score, the set of probabilities increases the chance to be selected as a "parent" for the next generation of new sets of probabilities.
In the end, the genetic algorithm retrieved the "genotype" of the stochastic CA with the best fitness score, composed of
the eight probabilities $p_i$ ($i=0,\dots,7$) for each of the eight 3-tuple 
$(c_{i-1,t},c_{i,t},c_{i+1,t})$.

The fitness function is based on the fitness measures of the avalanche size and duration with a 
power-law function and is heuristically defined as \citep{pontes2020neuro}
\begin{equation}
\label{eq8}
S_{temp} = (R^2)^2 
   + D^2   
   + B 
   + U,
\end{equation}
\begin{equation}
\label{eq:final_fit}
S = \begin{cases}
S_{temp}+L, &S_{temp}>3.5\\
S_{temp}, &\text{otherwise.}
\end{cases}
\end{equation}
where
$R^2$ is the coefficient of determination of complete linear fitting \citep{wright1921correlation},
$D$ is the normalized coefficient of the Kolmo\-gorov-Smirnov (KS) statistic \citep{clauset2009power},
$B$ is the percentage of non-zero bins with size one in the avalanche histograms,
$U$ is the percentage of unique states through time, and
$L$ is the normalized log-likelihood ratio of the comparison between the power-law model and the exponential model for estimating the avalanche distributions \citep{clauset2009power}. These fitness function objectives are normalized to the range $[0,1]$ if necessary and the genetic algorithm is applied to indirectly maximize them through the fitness function. The squared values in Eq.~\eqref{eq8}, $R^2$ and $D$, are the most important ones for the fitness function and were empirically chosen. In Eq.~\eqref{eq:final_fit}, the adjusted log-likelihood ratio $L$ is only calculated if $S_{temp}>3.5$ because this is a computationally intensive process; and if $L$ is not trustworthy ($p$-value of the ratio is greater or equal to 0.1), then this measurement is ignored (set as zero). In the end, we obtain the fitness score $S$.

\subsection{Adapted Model with Stochastic Transition Rates}

To test the robustness of the stochastic CA for remaining in criticality, Gaussian noise with a varying standard deviation $\sigma$ is applied to affect the probabilities $p_i$ of the CA for every time step. Since the Gaussian noise will make the perturbed probabilities pass the valid range between zero and one, normalization to the Gaussian mean (original probability) is used. Thus, the equation for the normalization is
\begin{equation}
\label{eq:norm_mean}
\tilde{\mu}_i=\log{\left ( \frac{p_i}{1-p_i} \right )}.
\end{equation}

With the normalized mean $\tilde{\mu}_i$, we sample $x_i$ from a Gaussian distribution, such as
\begin{equation}
\label{eq:gaussian_sample}
x_i \sim \mathcal{N}(\tilde{\mu}_i,\,\sigma).
\end{equation}
The random variable $x_i$ is sampled each time step, preserving the definition of CA, i.e. at each time step all cells follow the same (probabilistic) updating rule.
To make $x_i$ a valid perturbed probability $\tilde{p}_i$, the sigmoid function is applied to it, then
\begin{equation}
\label{eq:norm_prob}
\tilde{p}_i = \frac{1}{1 + e^{-x_i}}.
\end{equation}
While there are other possible choices, this choice maps a Gaussian distributed variable into a sigmoid function between $0$ and $1$ which for the average of $x_i$, $\tilde{\mu_i}$ retrieves the 
initial value of $p_i$.

Having settled this, our research question can be reformulated as follows: for $\sigma=0$ we have exactly the optimal solution, i.e.~a power-law of the size of clusters, their duration, and their area, then, by increasing $\sigma >0$ one will eventually destroy the power-laws observed for the optimal solution, pushing the system away from criticality. So, is there a transition from critical to non-critical stage tuning the $\sigma$?
We note that the maximum value of $\sigma$ for which criticality is still observed can be thought of as a measure of the robustness of the optimal (critical) stage.

\section{Results}
\label{sec:results}

\begin{figure*}[t]
\centering
\includegraphics[width=0.91\textwidth]{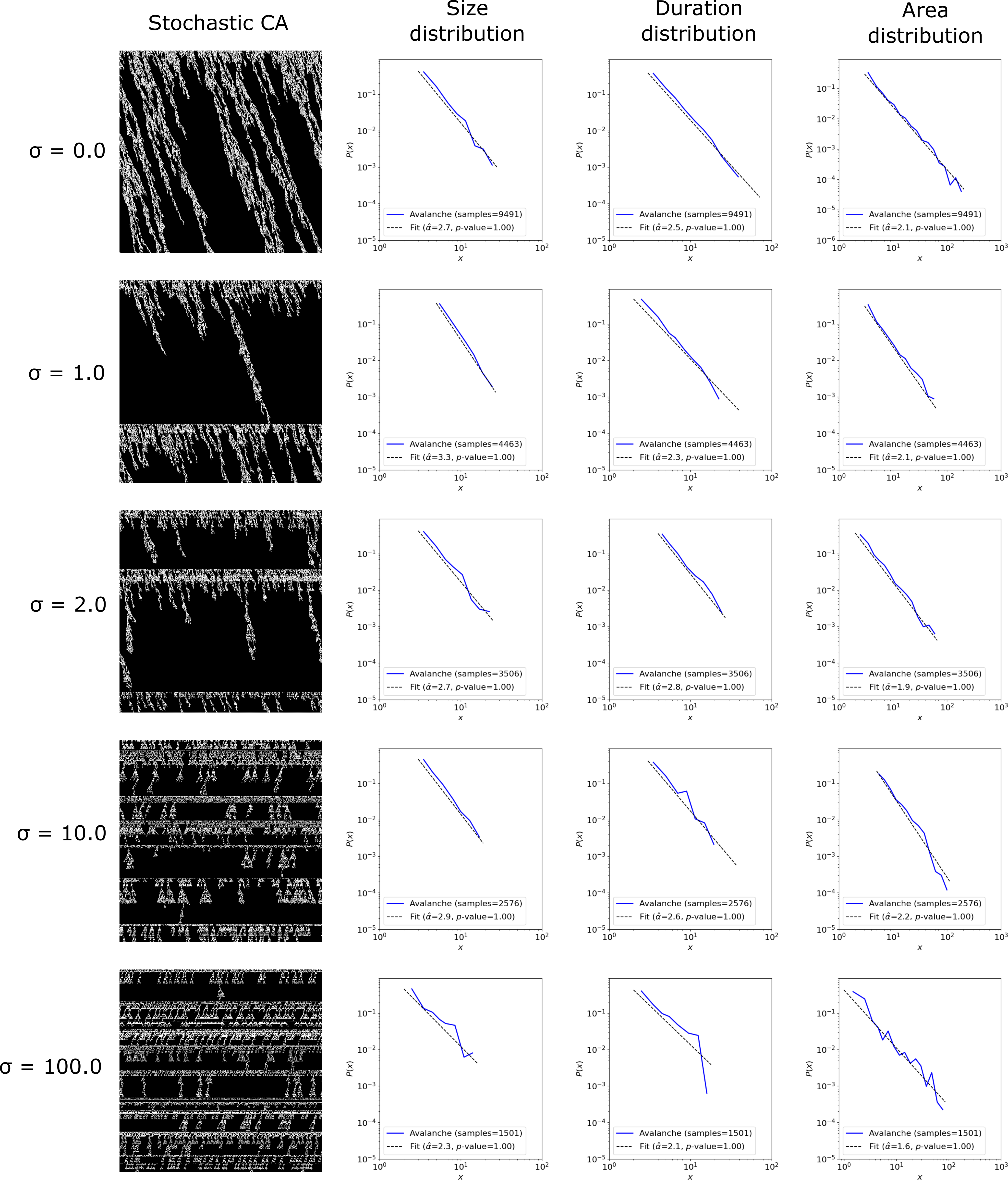}
\caption{Stochastic CAs and their avalanche distributions of the cells with state 1 (black cells) while affected by Gaussian noise. Here the CAs are produced with $N=1,000$ lattice squares and $T=1,000$ time steps. Large avalanches that happen only one time are ignored in these distributions.
The dashed lines indicate the power-law fit, with a slope defined as $\hat{\alpha}$ and a goodness-of-fit given by the $p$-value (see text).} 
\label{fig:pertubation}
\end{figure*}

For producing the perturbed stochastic CAs and the distribution of their area, size and duration; we repeat the simulations for the values of the standard deviation $\sigma$ from 0.1 to 2.0 in step size of 0.1, and also for values of 5, 10, 20, 50, 100, 200, 500, and 1,000. For each of those values of $\sigma$, we perform 1,000 CA simulations with a uniform random initialization. This is presented for some values of $\sigma$ in Fig.~\ref{fig:pertubation}. Its first row shows an illustrative realization of the optimal stochastic CA with the probabilities in Tab.~\ref{tab01}, showing the distribution of avalanche size, duration and area. A distribution is plotted as the avalanche measurement $x$ with its occurrence probability $P(x)$. The power-law fit is indicated with dashed lines and we can see that a power-law distribution fits well the empirical histogram. The power-law fit is also indicated by its estimated slope $\hat{\alpha}$ and goodness-of-fit given by the $p$-value. This $p$-value is introduced by \citet{clauset2009power}, stating that a $p$-value greater than 0.1 indicates a valid power-law fit.

\begin{figure*}[t]
\centering
\includegraphics[width=0.91\textwidth]{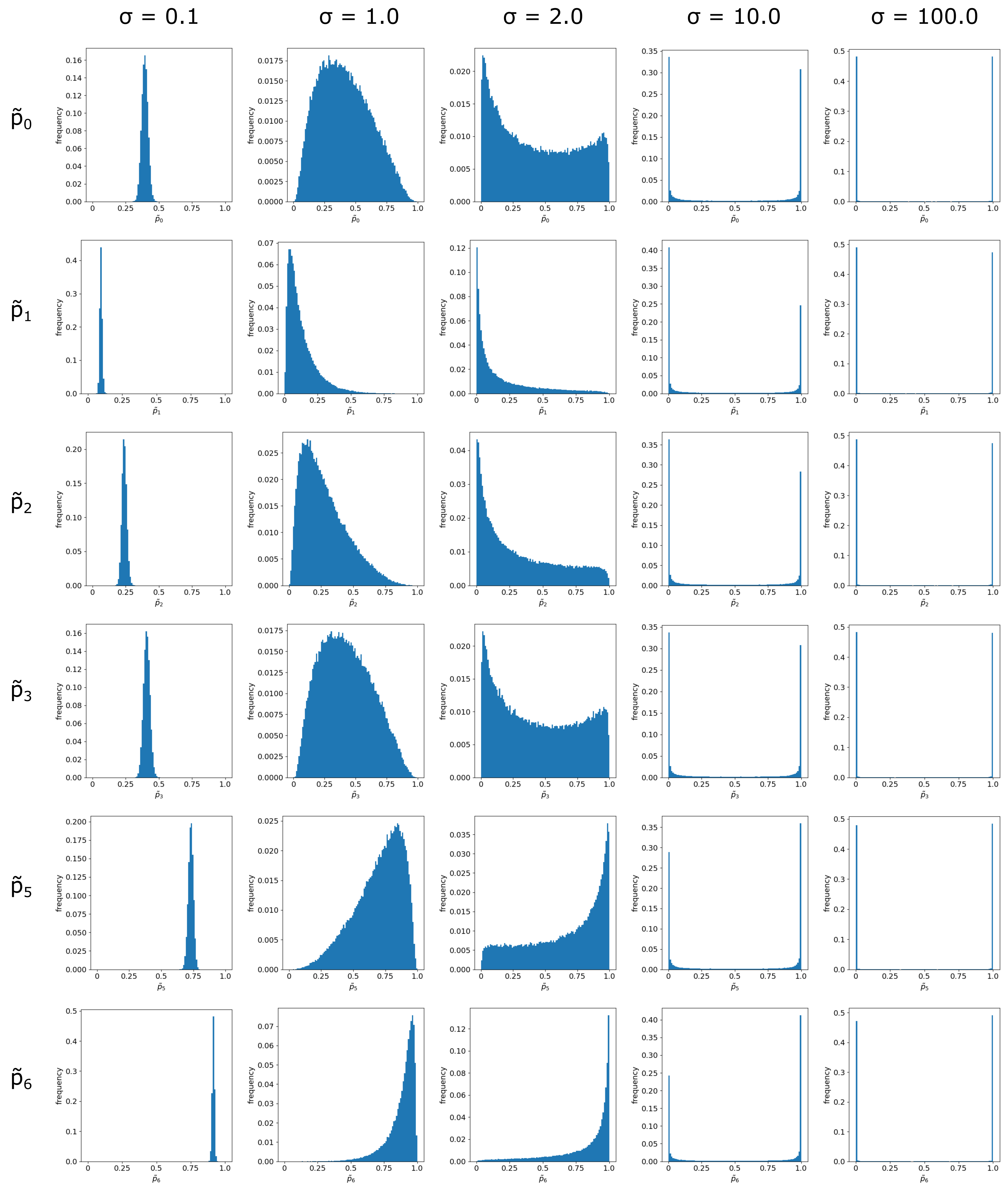}
\caption{Histogram of the valid perturbed probability $\tilde{p}_i$ for different values of $\sigma$. The probabilities $\tilde{p}_4$ and $\tilde{p}_7$ are not shown because they are always 0 and 1, respectively, independent of the value of $\sigma$. Each histogram was generated with 100,000 samples.}
\label{fig:histogram}
\end{figure*}

Fig.~\ref{fig:histogram} presents the histograms of the valid perturbed probability $\tilde{p}_i$ calculated through Eq.~\eqref{eq:norm_prob} for some values of $\sigma$. When $\sigma=2.0$, it is noticeable that the original Gaussian distribution becomes distorted. By increasing $\sigma$ even more, the distribution of the valid perturbed probability $\tilde{p}_i$ becomes concentrated in 0 and 1.

\begin{figure*}[tb]
\centering
\subfloat[]{\label{fig:number_ava_a}\includegraphics[width=0.45\textwidth]{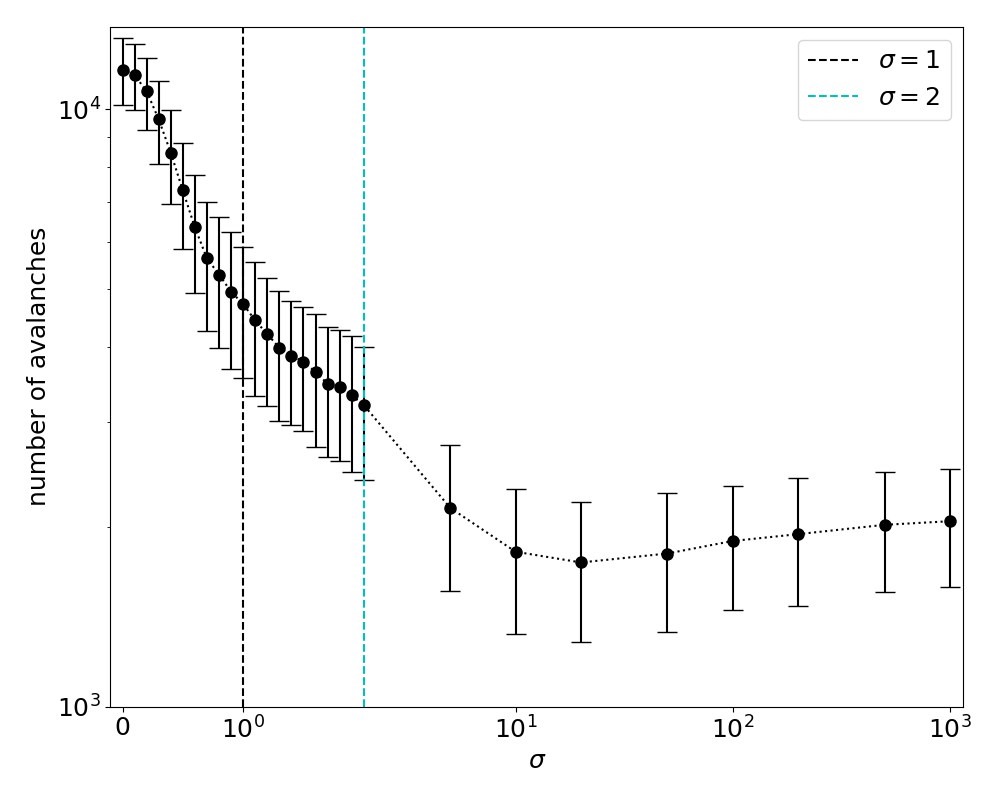}}
\subfloat[]{\label{fig:number_ava_b}\includegraphics[width=0.45\textwidth]{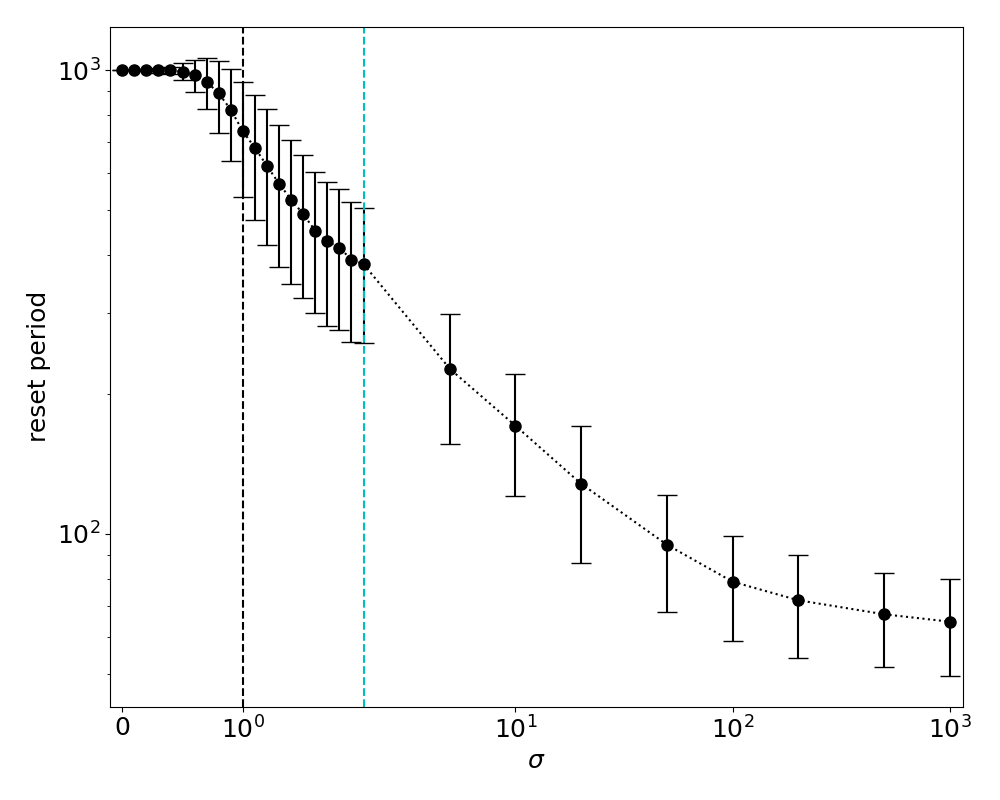}}
\caption{%
    Mean and standard deviation of 
    \protect\subref{fig:number_ava_a} the number of avalanches and
    \protect\subref{fig:number_ava_b} reset period.
}
\label{fig:number_ava}
\end{figure*}

\begin{figure}[tb]
\centering
\includegraphics[width=0.45\textwidth]{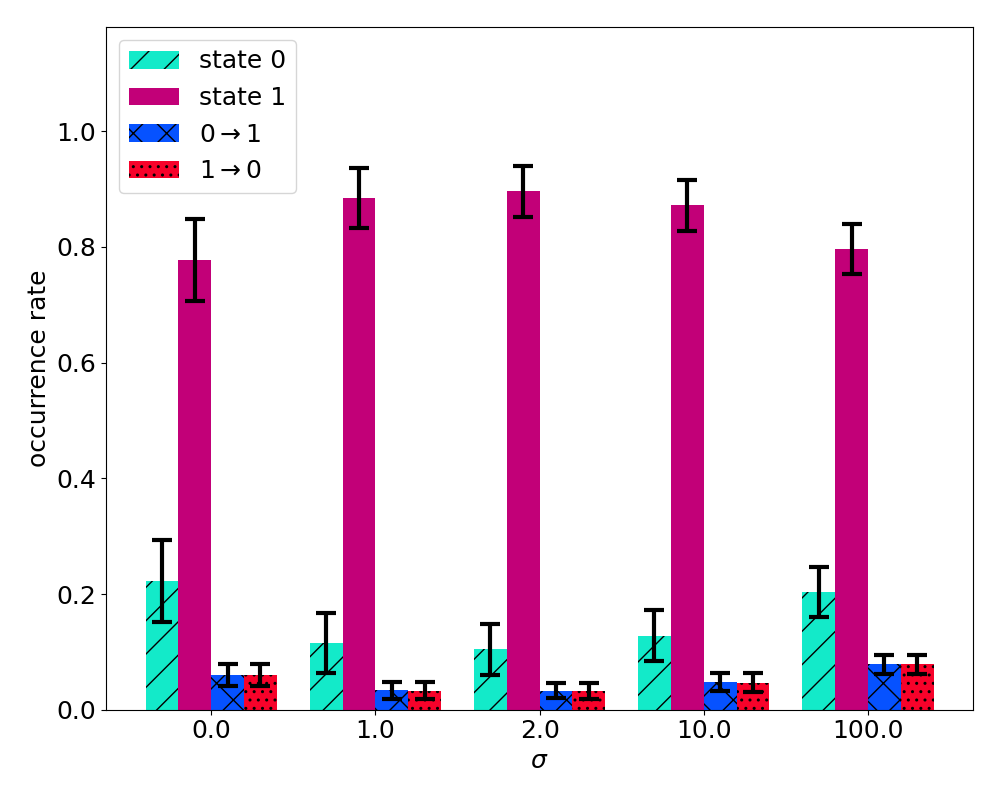}%
\caption{
Occurrence rates in the cells and their standard deviations of state 0 and 1, and the transitions between them that are from 0 to 1 ($0\rightarrow 1$) and from 1 to 0 ($1\rightarrow 0$).
}
\label{fig:states}
\end{figure}

Notice that, in case the strong quiescent state fully occupies the CA, its cells' states are re-initialized. 
For larger values of $\sigma$, since the CA reaches the inactivity state quicker than the optimal and unperturbed CA, the re-initialization happens more frequently.
In Fig.~\ref{fig:number_ava}, we plot the number of avalanches as well as the number of iterations before a re-initialization caused by all cells being in the quiescent state.
As one sees in this figure, due to a more frequent re-initialization the average avalanche size, duration and area is reduced. In Fig.~\ref{fig:pertubation}, this is not noticeable in the avalanche distributions because the large avalanches that happen just one time during the simulations are ignored. In those samples of the simulations, all goodness-of-fit $p$-value remains 1.0 as for the unperturbed CA.
The standard deviation $\sigma$ of the Gaussian noise to the probabilities strongly reduces the number of avalanches, especially from $\sigma$ between $0.3$ and $10.0$. After that, the number of avalanches stabilizes around 2,000. This is because the avalanches became so short that they reach the resting state with all cells in the strong quiescent state much faster, then re-initializing the CA more often and having a similar number of avalanches. Such a reduction in the number of avalanches of state 1 is also due to the decrease in the occurrence of state 0 because the avalanches of state 1 need to be surrounded by cells with state 0 or boundary. We can perceive that the state 0 patterns became not only shorter but also thinner. To confirm this behavior and to have a sense of what happens at individual cell scale, Fig.~\ref{fig:states} shows the occurrence rates and their standard deviation of state 0 and 1, and the transitions between these two states in both directions, from 0 to 1 ($0\rightarrow 1$) and from 1 to 0 ($1\rightarrow 0$). The occurrence rate of state 1 tends to increase until $\sigma=2.0$ because it is the strong quiescent state. The occurrence rate of state 0 and the transitions are inversely or directly proportional to the rate of state 1 because they are interdependent. However, for $\sigma>2.0$, the re-initialization of the CA counterbalances the trend of changes in those rates, even though it does not occur with the number of avalanches.

\begin{figure*}[tb]
\centering
\subfloat[Estimated slope $\hat{\alpha}$]{\label{fig:alpha}\includegraphics[width=0.33\textwidth]{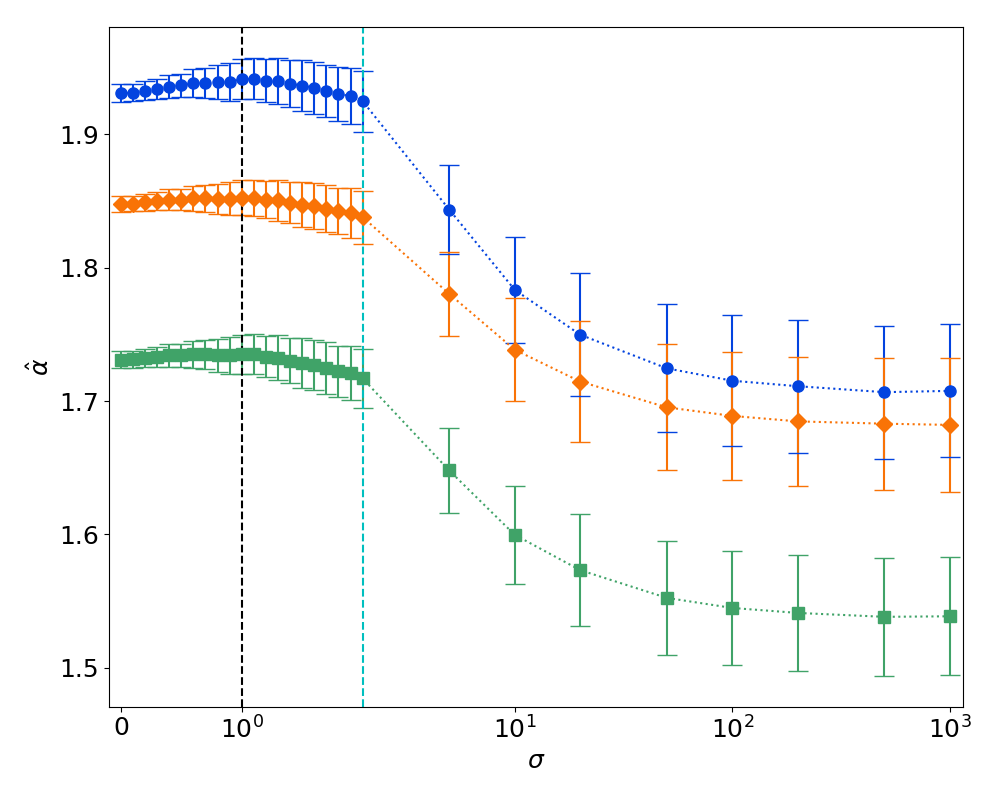}}%
\hfill
\subfloat[Kullback-Leibler divergence]{\label{fig:kl}\includegraphics[width=0.33\textwidth]{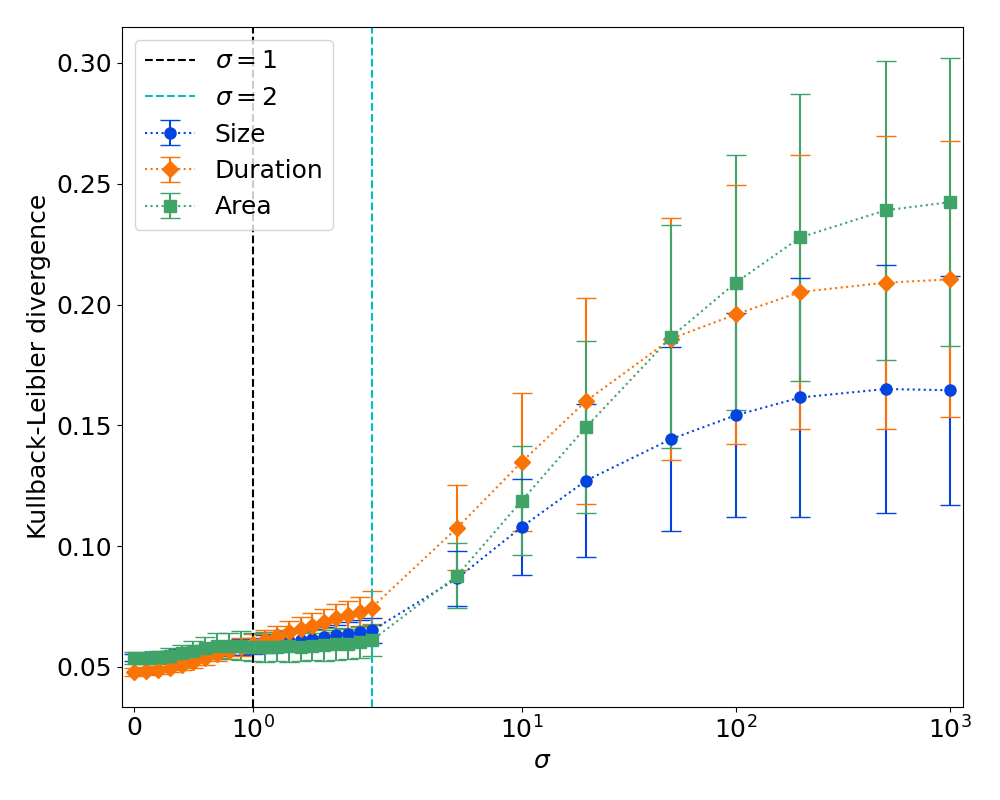}}%
\hfill
\subfloat[Kolmogorov-Smirnov statistic]{\label{fig:ks}\includegraphics[width=0.33\textwidth]{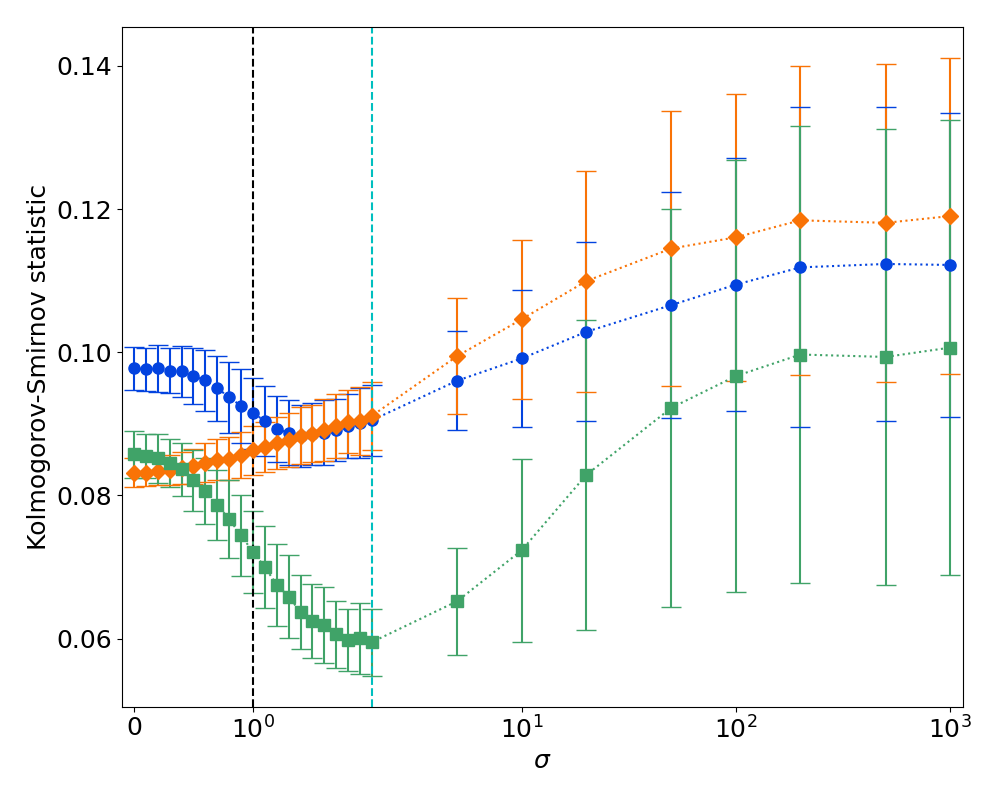}}%
\caption{
Mean and standard deviation of the measurements of the power-law estimation for the avalanche distributions of state 1.
}
\label{fig:error}
\end{figure*}

\begin{figure*}[tb]
\centering
\subfloat[Kullback-Leibler divergence]{\label{fig:kl_baseline}\includegraphics[width=0.36\textwidth]{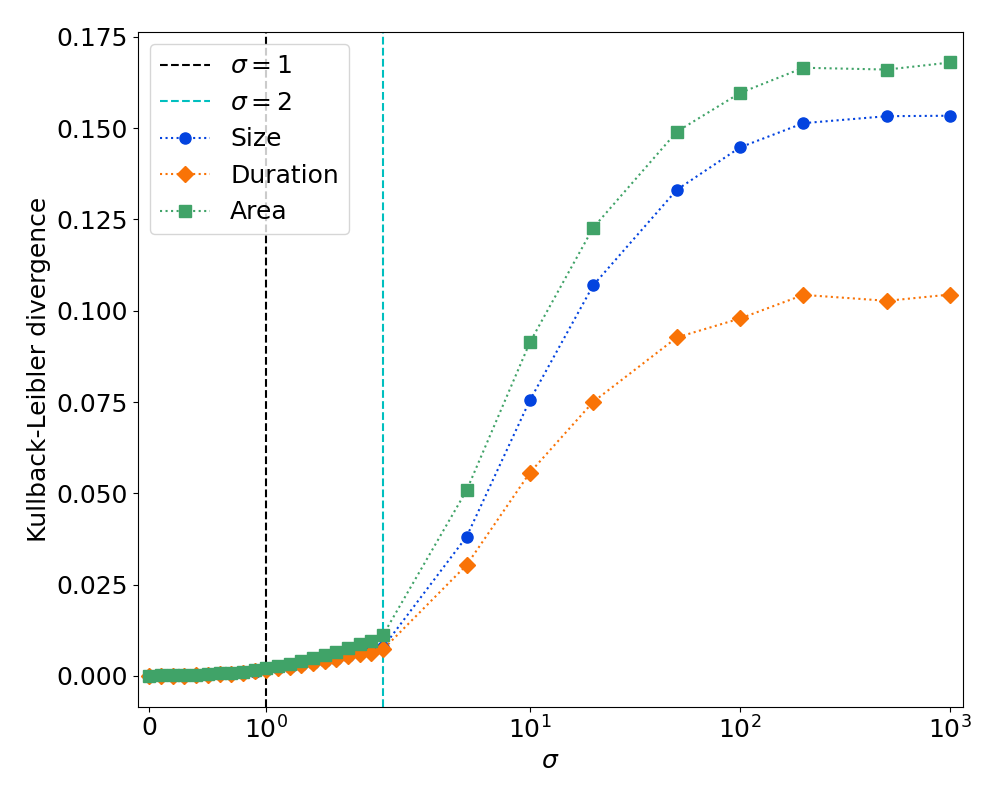}}%
\hspace{0.5cm}
\subfloat[Kolmogorov-Smirnov statistic]{\label{fig:ks_baseline}\includegraphics[width=0.36\textwidth]{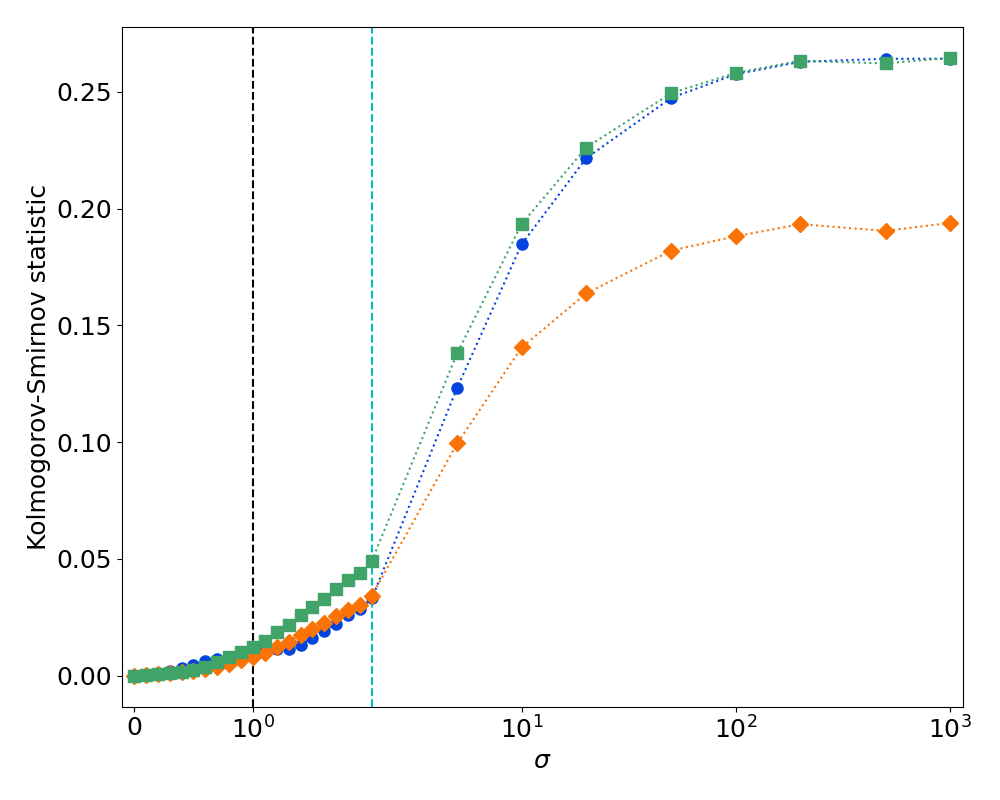}}%
\caption{
Comparison between the empirical distributions of each $\sigma$ and the empirical distribution with $\sigma=0$ used as a baseline.
}
\label{fig:baseline}
\end{figure*}

For the systematic evaluation of the robustness, we analyze the distributions of duration, size and area of the avalanches, by fitting to them a power-law and estimate their slope $\hat{\alpha}$,
as well as their uncertainty through a Kolmo\-gorov-Smirnov test and the corresponding Kull\-back-Leibler (KL) divergence \citep{kullback1951information,mackay2003information}.

The Kullback-Leibler divergence is defined as:
\begin{equation}
    D_{KL} = \sum_x P(x)log \left( \frac{P(x)}{Q(x)} \right)
\end{equation}
where $P(x)$ is the probability distribution of the empirical data and $Q(x)$ is the probability distribution function of the estimated power-law, and the supremum measuring the KS test is given by
\begin{equation}
    D_{KS} = \sup_x|P_{C}(x)-Q_{C}(x)| 
\end{equation}
where $P_C(x)$ is the cumulative distribution of $P(x)$ and $Q_C(x)$ is the cumulative distribution function of $Q(x)$.

Results are shown in Fig.~\ref{fig:error}: the slope is plotted in Fig.~\ref{fig:alpha}, while
the KL divergence is plotted in Fig.~\ref{fig:kl} and the KS test (supremum) statistic is shown in Fig.~\ref{fig:ks}.
Clearly, the slopes are almost constant for all distributions - size, duration and area - till a value of $\sigma\gtrsim 1$. Above $\sigma=2$, the estimated slopes reduce drastically, indicating 
the deviation from critical behavior.

The KL divergence $D_{KL}$ presents a constant (small) value for $\sigma\lesssim 1$, indicating a constant small error associated with the power-law fit. Beyond $\sigma=2$ this error increases considerably.
The KS statistic $D_{KS}$ shows a more fluctuating behavior. Indeed, counter-intuitively, before reaching $\sigma=1$, the supremum decreases to values smaller than the ones observed for $\sigma<1$.
This may be due to the occurrence of many small avalanches and one single large avalanche.

A comparison between the avalanche distributions of each $\sigma$ and the avalanche distributions of the stochastic CA with $\sigma=0$ is shown in Fig.~\ref{fig:baseline}. Therefore, avalanche distributions with $\sigma=0$ are used as a baseline for KL divergence (Fig.~\ref{fig:kl_baseline}) and KS statistic (Fig.~\ref{fig:ks_baseline}). This comparison indicates how the dynamics changes when additional noise is applied, and it confirms our observations from the KL divergences with their power-law fits.

\section{Discussion}
\label{sec:conclusions}

The optimal stochastic CA without perturbations ($\sigma=0$) and probabilities shown in Tab.~\ref{tab01} can maintain its critical behavior even when perturbed ($\sigma>0$), but only up to a certain point. By analyzing Figs.~\ref{fig:error} and \ref{fig:baseline}, we can conclude that $\sigma=1$ is the breaking point between behaving and not behaving similarly to the unperturbed one. This can be noticed especially in Fig.~\ref{fig:alpha}. The estimated slopes $\hat{\alpha}$ for size, duration and area remain almost unvaried until $\sigma=1$, while their standard deviations slowly increase as a result of the fluctuations in the values of the probabilities. Even though the behavior is maintained with respect to the estimated slopes $\hat{\alpha}$, KS statistic and KL divergence; Fig.~\ref{fig:number_ava_a} shows that the number of avalanches starts to decrease from approximately 12,000 to around 4,000 in the range $\sigma=[0, 1]$. This cannot be seen in the reset period (Fig.~\ref{fig:number_ava_b}) because the number of time steps simulated is 1,000, making it the maximum reset period possible.

Neuronal stochastic variability \citep{mcdonnell2016neuronal} happens all over the brain and on all scales. Since the optimal stochastic CA investigated in this work presented robustness up to perturbations with $\sigma=1$, biological neural networks may also allow for similar robustness to perturbations. Therefore, both systems may preserve criticality even in noisy conditions. Because self-organized criticality is possibly one of the main factors for the emergence of intelligence in the human brain \citep{heiney2021criticality}, the evaluation of robustness to stochastic variability can be an important measurement for indicating the presence of intelligence in artificial systems.

In future work, we plan to evaluate the performance of the optimal stochastic CA under perturbations in a machine learning framework called reservoir computing \citep{lukovsevivcius2009reservoir,yilmaz2014reservoir,nichele2017deep}. Reservoir computing is a biologically-plausible neural model inspired by the functioning of cortical microcircuits \citep{maass2002real}. There is indeed evidence that reservoir computing achieves better performances when it produces critical dynamics \citep{suarez2021learning,glover2021dynamical,boedecker2012information}. This benchmark would inform how the artificial intelligence system can maintain its accuracy while increasing the probability noise. Therefore, such investigation may confirm that this robustness is also essential for future artificial intelligence systems, and in particular inform the realization of novel brain-inspired neuromorphic hardware.

\section*{Acknowledgements}
This work was supported by the Norwegian Research Council SOCRATES project (grant number 270961).

\FloatBarrier
\bibliographystyle{unsrtnat}
\bibliography{references}

\end{document}